\documentclass[11pt ]{article}
\usepackage{amsmath}
\usepackage{graphicx}
\usepackage{hyperref}
\usepackage{lmodern}
\usepackage{amssymb}
\usepackage{booktabs}
\usepackage{tikz}
\usetikzlibrary{patterns}

\newcommand{\be}{\begin{equation}}
\newcommand{\ee}{\end{equation}}
\newcommand{\beqa}{\begin{eqnarray}}
\newcommand{\eeqa}{\end{eqnarray}}

\renewcommand{\theequation}
{\arabic{section}.\arabic{equation}}

\makeatletter
\def\eqnarray{ \stepcounter{equation} \let\@currentlabel=\theequation
 \global\@eqnswtrue
 \global\@eqcnt\z@
 \tabskip\@centering
 \let\\=\@eqncr
 $$\halign to \displaywidth\bgroup\@eqnsel\hskip\@centering
 $\displaystyle\tabskip\z@{##}$&\global\@eqcnt\@ne
 \hfil$\displaystyle{{}##{}}$\hfil
 &\global\@eqcnt\tw@$\displaystyle\tabskip\z@{##}$\hfil
 \tabskip\@centering&\llap{##}\tabskip\z@\cr}
\makeatother

\makeatletter
\def\@arrayacol{\edef\@preamble{\@preamble \hskip .5\arraycolsep}}
\def\array{\let\@acol\@arrayacol \let\@classz\@arrayclassz
\let\@classiv\@arrayclassiv \let\\\@arraycr\def\@halignto{}\@tabarray}
\makeatother

\makeatletter
\newcounter{subeqncnt}
\def\thesubeqncnt{\alph{subeqncnt}}
\def\subequations{\begingroup%
   \stepcounter{equation}\edef\@tempa{\theequation}%
   \let\c@equation\c@subeqncnt\c@subeqncnt\z@
   \edef\theequation{\@tempa\noexpand\thesubeqncnt}}

\makeatother

\newcommand{\nn}{\nonumber}

\def\CD {{\cal D}}

\def\CM {{\cal M}}

\def\CO {{\cal O}}

\begin{document}

\setlength{\baselineskip}{6mm}
\begin{titlepage}
\begin{flushright}

{\tt NRCPS-HE-10-2018} \\

\end{flushright}

\begin{center}
{\Large ~\\{\it  Lecture on Quantum Gravity with  Perimeter  Action  \\
 \vspace{0,5cm}
 and \\
  \vspace{0,5cm}
Gravitational Singularities\footnote{${}$ Lecture presented at the Corfu Summer Institute 2017 "School and Workshops on Elementary Particle Physics and Gravity",
                 2-28 September 2017, Corfu, Greece and at the Workshops on Elementary Particle Physics and Gravity", 2-28 September 2017, Corfu, Greece and COST Action MP1405   "Quantum Structure of Spacetime"
III.  Annual Workshop: Quantum Spacetime '18, 19 - 23 February 2018,  Sofia, Bulgaria. }   
}

}

\vspace{2.5cm}

 {\sl  George Savvidy

 \bigskip
 \centerline{${}$ \sl Institute of Nuclear and Particle Physics}
\centerline{${}$ \sl Demokritos National Research Center, Ag. Paraskevi,  Athens, Greece}
\bigskip

}
\end{center}
\vspace{30pt}

\centerline{{\bf Abstract}}
Motivated by  quantum-mechanical considerations we earlier suggested  an alternative action for discretised quantum gravity which measures the perimeter of the space-time and has a dimension of length.  It is the so called {\it perimeter action}, since it is a "square root" of the   {\it area  action} in gravity and has a new constant of dimension one in front. The physical reason to introduce the perimeter/linear action was to suppress singular configurations "spikes" in the quantum-mechanical integral over geometries. Here we shall consider the continuous limit of the discretised perimeter/linear action.   We shall demonstrate that in the modified theory during the time evolution of a large massive star, when a star undergoes a collapse and develops an event horizon which confines the light,  a smaller space-time region will be created behind the event horizon which is unreachable by test particles. These regions are located in the places where a standard theory of gravity has singularities.  We are confronted  here with a drastically new concept that during the time evolution of a massive star a space-time region is created which is excluded from the physical scene, being physically unreachable by test particles or observables. If this concept is accepted, then it seems plausible that the gravitational singularities are  excluded from the modified theory.

\vspace{12pt}

\noindent

\end{titlepage}

\section{ \it Area  Action Versus  Perimeter Action}

Unification of gravity with other fundamental forces within the superstring theory stimulated the interest to the theory of quantum gravity and to physics at Planck scale 
\cite{sacharov,Buchdahl,Starobinsky:1980te,Adler:1982ri,Gasperini:1992em}. In particular, string theory predicts modification of the gravitational action 
at Planck scale with additional high derivative terms. This allows to ask fundamental questions concerning physics at Planck scale referring to these effective actions  and, in particular, one can try to understand how they influence the gravitational singularities 
\cite{Penrose:1964wq,Christodoulou,Hawking1,Hawking2,Hawking3,HawkingPenrose,Robertson,Raychaudhure,Komar,Markov:1982ed,anini,Brandenberger:1988aj,Alvarez:1984ee,Baierlein:1962zz,Brandenberger:1993ef,Chamseddine:2016ktu,Chamseddine:2016uef,Deser:1998rj}.

It is appealing to extend  this approach to different modifications of classical gravity which follow from  string theory and also to develop an alternative approach which is based on new geometrical principles \cite{Savvidy:1995mr,Ambjorn:1996kk,Savvidy:1997qf}\footnote{See also the references \cite{Ambjorn:1985az,Ambartsumian:1992pz, Ambjorn:1997ub,Savvidy:2015ina}.}. 
This approach to quantum gravity is based on the idea that the quantum mechanical amplitudes should be proportional to the "linear size" of the geometrical fluctuations of the space time manifold. 
This principle will allow to extend the notion of the Feynman integral over paths to an integral over space-time manifolds so that when {\it a  manifold collapses into a single world line} the corresponding quantum-mechanical amplitude becomes proportional to the {\it length of the world line}. In other words, in this limit the gravitational action should reduce to the relativistic particle action which is equal to the length of the world line and  measures it in $cm$ \cite{Savvidy:2015ina}.   

The legitimate question to be asked is why to consider alternative geometrical principles? The reason is that when the action has a dimension larger than one,  that is, the action has 
dimension $cm^d$, where $d > 1$,  then the  geometrical fluctuations of lower dimension will grow 
uncontrollably on a space-time manifold.  This happens because the action is "blind"  toward measuring the low dimensional fluctuations \cite{Ambjorn:1985az,Ambartsumian:1992pz, Ambjorn:1997ub,Savvidy:2015ina}.  
Indeed, let us consider a discretised two-dimensional world sheet surface and a theory in which the action is equal to the area of the surfaces. The Feynman integral here is an integral over all vertices of the triangulated surface.
The fluctuations will grow on a surface in the form of tine spikes, because spikes have zero  area and will be created with a large amplitude of order one.  The benefit of introducing a perimeter/linear action is that it suppresses singular configurations in the form of "spikes" in the quantum-mechanical integral over geometries, thus suppressing any fluctuation of lower dimensionality.

One can  demonstrate this phenomenon on a beautiful example from Integral Geometry \cite{chern,ambartzumian,santalo}. Let us consider a random triangle $ABC$ on a two-dimensional plane which is created by three randomly distributed  vertices $ \vec{x}_1,\vec{x}_2,\vec{x}_3 $ \cite{ambartzumian}.  The measure which is invariant under the isometries of the Euclidean plane is (see Fig.\ref{fig0}) $$d\mu =d^2 x_1d^2 x_2 d^2 x_3,$$
and we shall consider the partition functions when the action is proportional to the 
area of the triangle $S$ or to its perimeter $L$.  
Using the triangle angles $\alpha_{1,2,3}$  one can represent the measure $d\mu$ in the  form 
$
d\mu = d^2 x_1 d\phi ~ \rho_2 \rho_3 d \rho_2 d \rho_3 d\alpha_1,
$
where $d^2 x_2 = \rho_2   d \rho_2  d\phi_2$, ~$ d^2 x_3 = \rho_3   d \rho_3  d\phi_3$,~
$\phi_3 - \phi_2 = \alpha_1$ and $\phi_2 = \phi$ and then express it in the following form
\cite{ambartzumian}:
\begin{figure}
\begin{center}
\includegraphics[width=7cm]{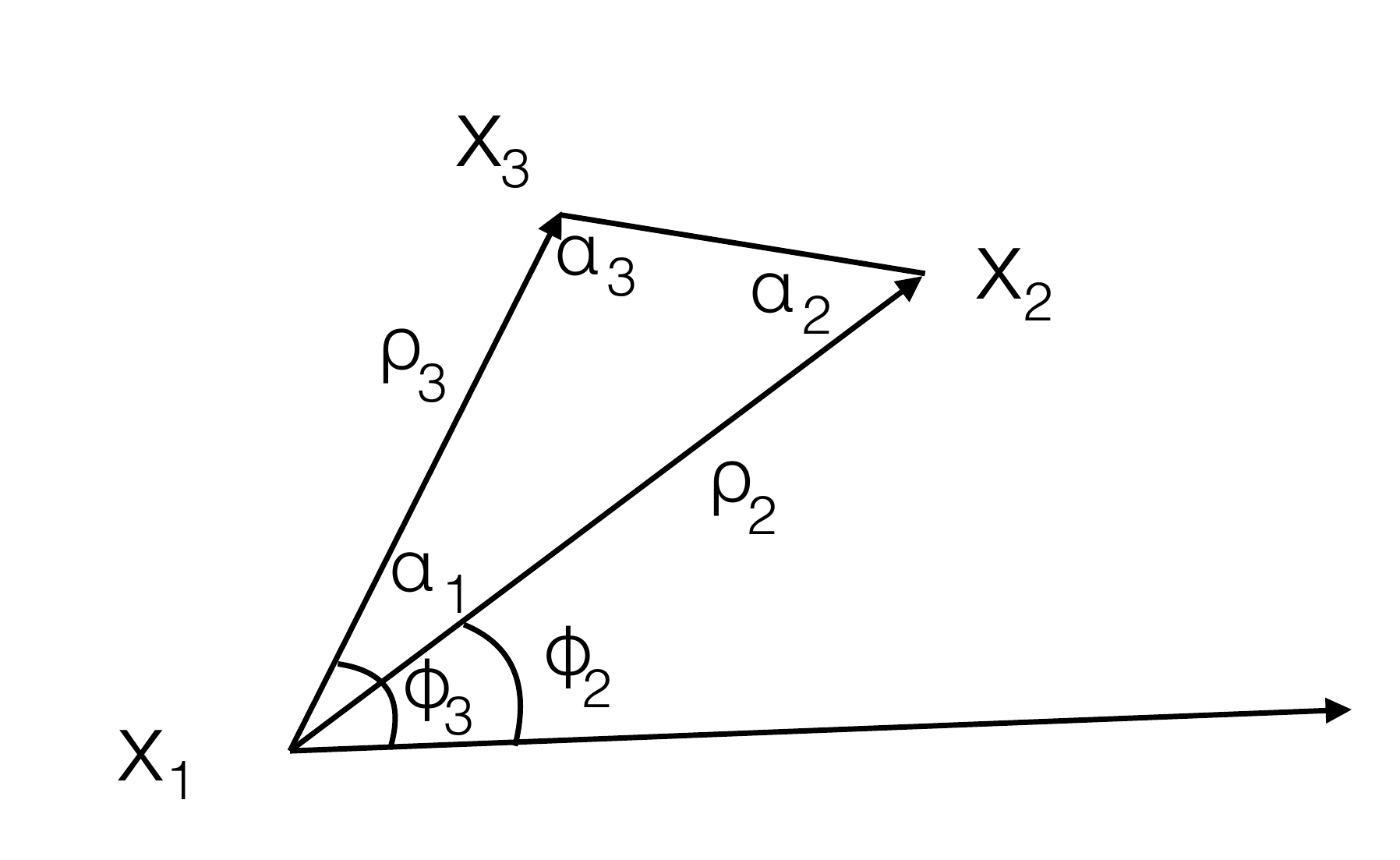}~~~~~~~~~~
\caption{The measure which is invariant under  isometries of the Euclidean plane is $d\mu =d^2 x_1d^2 x_2 d^2 x_3$.  It can be transformed into the form which is expressed in terms of geometrical characteristics of the triangle shape such 
as its area $S$ or perimeter $L$ and the angles $\alpha_1$, $\alpha_2$. The results are presented  by formulas (\ref{area1}) and (\ref{perimeter}). 
}
\label{fig0}
\end{center}
\end{figure}
\be\label{area1}
d\mu_S = d^2 x_1 d\phi ~~ S dS {d\alpha_1 d\alpha_2 \over \sin \alpha_1 \sin \alpha_2  \sin \alpha_3}
\ee
and as 
\be\label{perimeter}
d\mu_L = d^2 x_1 d\phi ~~ L^3 dL {\sin \alpha_1 \sin \alpha_2  \sin \alpha_3 \over (\sin \alpha_1 + \sin \alpha_2 + \sin \alpha_3)^4} d\alpha_1 d\alpha_2 .
\ee
The part of the measure $d^2 x_1 d\phi$ factorises,  it describes the translation and rotation of the triangle as a whole and therefore is irrelevant for our consideration. The rest of the measure allows to calculate the entropy, that is to answer the following question: how many of the randomly created triangles have the area $S$ or the perimeter $L$?  For that one should integrate the measure in (\ref{area1}) and (\ref{perimeter}) over the independent parameters defining a geometrical shape of the triangles:  $\alpha_1$ and $\alpha_2$ at fixed $S$ or $L$.  The integral is logarithmically diverging in the area case (\ref{area1}) and is finite in the perimeter case (\ref{perimeter}). Thus the results are 
different: there are infinitely many triangles of fixed area, because the triangle can be infinitely long, in the form of a spike.  In the case of the perimeter action (\ref{perimeter}) the integral is converging and is perfectly well defined.  This example illustrates why the perimeter action has an advantage to define a geometrical theory in which the spiky configurations are suppressed.

\begin{figure}
\begin{center}
\includegraphics[width=6cm]{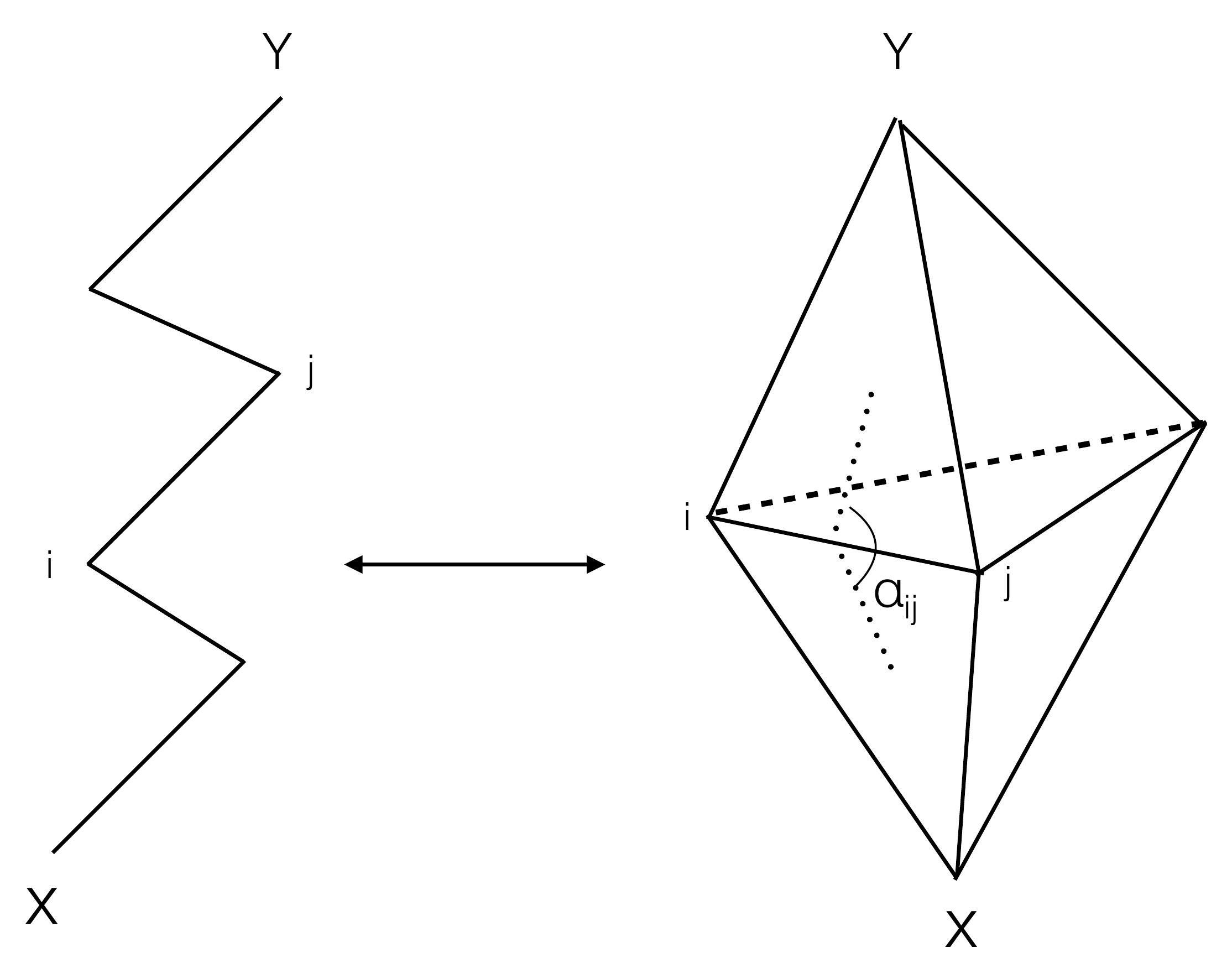}~~~~~~~~~~
\caption{
The discretised two-dimensional surface describing the propagation of a string from space-time point $X$ to $Y$.   The action (\ref{actionparticle}) allows to extend the notion of the Feynman integral over paths to an integral over space-time surfaces, so that when a two-dimensional surface degenerates into a single world line the quantum mechanical amplitude becomes proportional to the {\it length of the world line} :~~  $L= \sum_{<ij>} \lambda_{ij} ~~~ 
 \longleftrightarrow ~~~ L =  \sum_{<ij>} \lambda_{ij} \vert \pi -\alpha_{ij} \vert $. 
}
\label{fig1}
\end{center}
\end{figure}

In the case of a single triangle geometries it was clear what should be understood under its perimeter or of its linear size.  {\it The  question is how to measure  the perimeter/linear size of the high-dimensional manifolds} and, in particular, a triangulated two-dimensional world sheet surface in terms of $cm$, instead of the areas of its triangles. The invariant which characterises the linear size of the discretised  two-dimensional surface can be constructed  summing the lengths of its edges $l_{ij}$ multiplied by the deficit angle $\omega_{ij}=\vert \pi -\alpha_{ij} \vert$ on the corresponding edge $<ij>$  \cite{Savvidy:1995mr,Ambjorn:1996kk,Savvidy:1997qf,Ambjorn:1985az,Ambartsumian:1992pz, Ambjorn:1997ub,Savvidy:2015ina} (see Fig.\ref{fig1})
\be\label{actionparticle}
L= \sum_{<ij> \in \CM_2} \lambda_{ij} \vert \pi -\alpha_{ij} \vert .
\ee
The action measures the surface in terms of $cm$ and reduces to its length when the surface collapses into a single world line, as one can see on Fig.\ref{fig1}. The deficit angle in the above formula plays an important role, because otherwise the action will be ill defined and unbounded,
in particular, if one adds a flat edge with its dihedral angle $\alpha_{ij}=\pi $ it will not 
contribute into the action, only non-flat edges $\alpha_{ij} \neq \pi $ contribute into the sum over edges.

\begin{figure}
\begin{center}
\includegraphics[width=8cm]{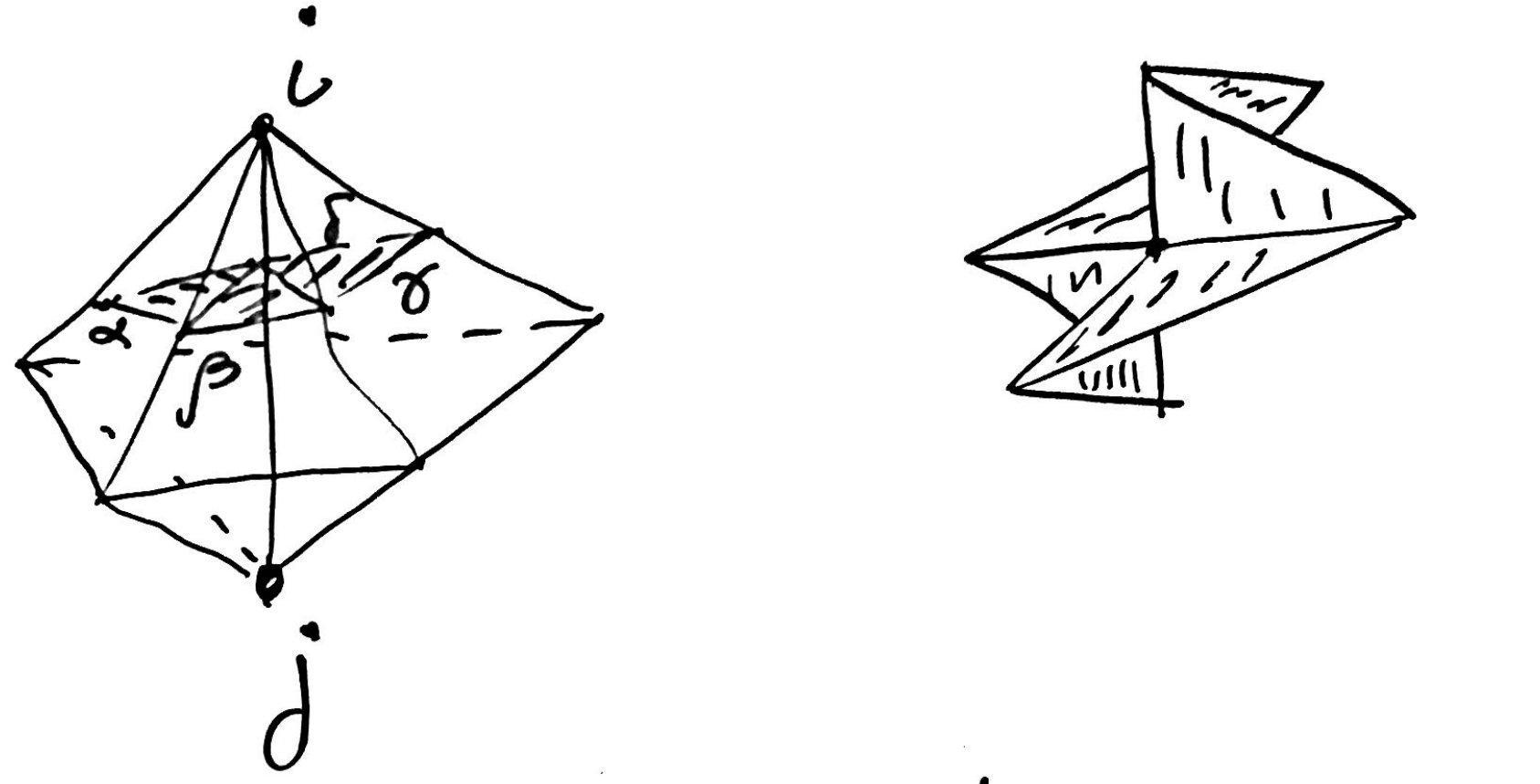}~~~~~~~~~~
\caption{
The example of discretised space-time manifold of the three-dimensional gravity.  In three dimensions  the Regge action (\ref{threegravity})  measures the linear size of the three-dimensional space-time as it does its continuous counterpart (\ref{threegravi}).  The Regge action (\ref{threegravity}) is a sum of  lengths of all edges of the three-dimensional simplex multiplied by the deficit angles on cones which appear in the intersection of the edges by the normal planes. On the right hand side one can see an example of such a cone. The deficit angle is equal to the curvature of the cone $\omega_{ij}= (2 \pi - \alpha -\beta-\gamma-\delta)_{ij}$.  If the cone is flat $(\alpha +\beta+\gamma+\delta)_{ij} = 2\pi$ then the deficit angle $\omega_{ij}$ is equal to zero and the edge does not contribute to the action.  
}
\label{fig2}
\end{center}
\end{figure}

Let us now turn to a three-dimensional space-time manifolds representing a discretised quantum gravity. In that case the  action was found by Regge and it  has the form \cite{Regge:1961px}:
\be\label{threegravity}
L = \sum_{<ij> \in \CM_3} l_{ij} ~\omega_{ij}, ~~~~~~~\omega_{ij}=  (2\pi -\alpha_{1} -...-\alpha_{n})_{ij}
\ee
where $ l_{ij}$ is the {\it length} of the edge $<ij>$ and $\omega_{ij}$ is the deficit angle on the edge (see Fig.\ref{fig2}).
This action is the discretised version of the Hilbert-Einstein (HE) action,  it has the dimension of length
\be\label{threegravi}
L = \int_{\CM_3} R \sqrt{-g} d^3x
\ee
and measures the linear size of the three-dimensional manifold in $cm$.

\begin{figure}
\begin{center}
\includegraphics[width=10cm]{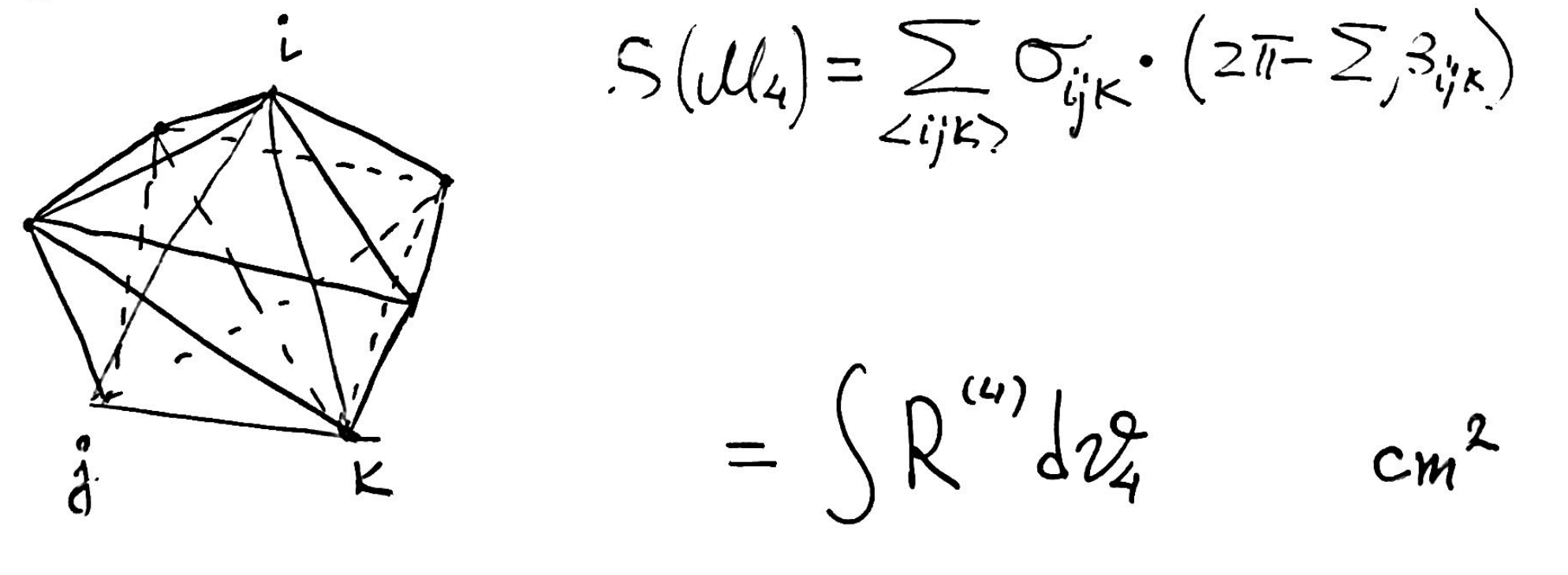}~~~~~~~~~~
\caption{
The example of the simplicial space-time manifold of four-dimensional gravity.   The Regge action (\ref{areafour}) is a sum of  all areas of the triangles multiplied by the deficit angles on a cone which appears in the intersection of the triangles by the normal planes.  The perimeter/linear action (\ref{lamb1}) in four-dimensions is constructed  by replacing the areas of the triangles $\sigma_{ijk}$ in (\ref{areafour}) by the perimeters $\lambda_{ijk}$, as we were advocated 
in the text (\ref{area1}), (\ref{perimeter}) and demonstrated  on Fig. \ref{fig0}. }
\label{fig3}
\end{center}
\end{figure}

Finally in four dimensions the HE action has dimension $cm^2$ and measures the 
area of the universe. The dimension of the measure $ [ \sqrt{-g} d^4x]$ is 
$cm^4$, the dimension of the scalar curvature  $[R] $ is ${1/ cm^2} $, thus the 
integral $\int R \sqrt{-g} d^4x$ has dimension $cm^2$ and measures  the area of the universe 
Fig.\ref{fig3}. In the discretised representation the Regge action is the sum of {\it areas} of the   triangles  multiplied by the corresponding deficit angles \cite{Regge:1961px}
\begin{equation}\label{areafour}
S= \sum_{<ijk> \in \CM_4} \sigma_{ijk} ~ \omega_{ijk}, 
\end{equation}
where $\sigma_{ijk}$ is the area of the triangle $<ijk>$ and $\omega_{ijk}$ is the deficit angle on the triangle $<ijk>$. Here as well, there is a cone which appears in the normal section of the triangle $<ijk>$ and  the deficit angle is equal to its curvature. The action represents the discretised version of the HE area action in four-dimensions: 
\be\label{standgravity}
S =  \int_{\CM_4} R \sqrt{-g} d^4x .
\ee
We have been arguing above (\ref{area1}), (\ref{perimeter}) that the area action is unable to measure the one-dimensional singular configurations appearing in the form of tiny  spikes and that the {\it linear functional}  similar to the Feynman path integral action for the relativistic particles can represent a desired solution.

This raises a question: Is it possible to construct an action which measures the linear size of  the four-dimensional manifold? It is not so difficult to construct an appropriate linear action
starting from the Regge action (\ref{areafour}) and taking instead of the triangle area  $\sigma_{ijk}$ its perimeter $\lambda_{ijk}$ \cite{Savvidy:1995mr,Ambjorn:1996kk,Savvidy:1997qf}
\begin{equation}
L = \sum_{<ijk> \in \CM_4}\lambda_{ijk} \cdot \omega_{ijk}.
\label{lamb1}
\end{equation}
The linear character of the  action (\ref{lamb1}) requires the 
existence of a new fundamental coupling constant $m_P$ of dimension $ 1/cm$.
It is natural to call this action  "perimeter/linear" or "gonihedric" because its definition contains the sum of products of the characteristic lengths and deficit angles.  
The action is well defined for discretised $4D$ manifolds and can also be derived by postulating
geometrical principles:  the linearity and the continuity of the action functional.

Indeed, the discretised  version of the linear action can be derived from: 
{{\bf $\alpha )$}} the coincidence of the perimeter/linear action 
with the Feynman path integral in the cases when a manifold  collapses to a single world line and {{\bf $\beta )$}} the continuity of the transition 
amplitudes under the manifold deformations \cite{Savvidy:1995mr,Ambjorn:1996kk,Savvidy:1997qf}. In accordance with $\alpha )$ the quantum mechanical amplitude should be proportional to the {\it length of the space-time manifold} and therefore  it must be proportional to the linear combination of the lengths of all edges of the 
discretised space-time  manifold
$
 \sum_{<i,j>} \lambda_{ij}\cdot \Theta_{ij}
$, 
where $\lambda_{ij}$ is the length of the edge between 
two vertices $<i>$ and $<j>$, summation is over all
edges $<i,j>$  and $\Theta_{ij}$ is an unknown angular factor, which can be defined through the continuity principle $\beta )$.  The deficit angel $\Theta_{ij}$ should  vanish in the cases when the triangulation around the edge $<i,j>$ is  $flat$, thus 
$\sum_{<i,j>}\lambda_{ij}\cdot \sum (2\pi - 
\sum \beta_{ijk}),$
where  $\beta_{ijk}$ are the angles on the cone which appear in the normal section of the edge $<ij>$. Combining terms belonging to a given triangle
$<ijk>$ we shall get a sum $\lambda_{ij}+\lambda_{jk}+\lambda_{ki}=
\lambda_{ijk}$ which is equal to the perimeter of the triangle $<ijk>$ 
and $\omega_{ijk}$ is the deficit angle on 
the triangle $<ijk>$ \cite{Savvidy:1995mr,Ambjorn:1996kk,Savvidy:1997qf}, thus recovering 
the action (\ref{lamb1}). 
Thus the principles of {\it linearity} and  {\it continuity}  allow to define the perimeter/linear action $L$ which can be considered as a "square root" of classical Regge area $S$ action (\ref{areafour}) \cite{Regge:1961px,wheeler}.

Comparing the linear  action $L$ in (\ref{lamb1}) and the Regge area action  $S$ in (\ref{areafour}) one should emphases  that there is a deep analogy between these expressions and the earlier example considered in the beginning of this section where we were arguing that the perimeter action (\ref{perimeter}) has advantage compared with area action   (\ref{area1}) since the integration over all geometries by the perimeter/linear action is well defined and nonsingular. Guided by this consideration that the perimeter/linear action has an advantage to suppress  singular quantum-mechanical amplitudes, it is desirable to find a continuous counterpart  of the perimeter/linear action (\ref{lamb1}). 
In  case of the Regge action it was proven that its continuous limit reduces to the HE action (\ref{standgravity}) and our aim here is to address the same question in the  case of the 
perimeter/linear action (\ref{lamb1}). 

\section{\it  Perimeter Action in Continuous Limit}

It is unknown to the author how to derive a continuous limit  of the linear action (\ref{lamb1}) in a unique way. In this circumstance we shall try to 
construct a possible linear action for a smooth space-time universe by using the available geometrical invariants.  Any expression 
which is quadratic in the curvature tensor and includes two derivatives could be a candidate for the linear action.  The invariants  we have chosen have the following form:
\beqa\label{lineargravity2}
 I_1= -{1\over 180}R_{\mu\nu\lambda\rho;\sigma} R^{\mu\nu\lambda\rho;\sigma} ,~~~~~~~~~~I_2=+{1\over 36} R_{\mu\nu\lambda\rho}  \Box  R^{\mu\nu\lambda\rho}~,  
\eeqa
and we shall consider a linear combination of the above expressions\footnote{The general form of the action is presented in the Appendix.}:
 \beqa\label{lineargravity}
 && L=  -M c
  \int_{\CM_4} {3  \over 8 \pi} (1-\gamma) \sqrt{I_1 +\gamma I_2 }~\sqrt{-g }  d^4x,
\eeqa
where we introduced the corresponding mass parameter $M$ and the dimensionless parameter $\gamma$. The dimension of the  invariant  $[\sqrt{I_1 + \gamma I_2 }] $ is ${1/ cm^3} $, thus the invariant $L$ has the dimension of $cm$ and measures  the linear "size" of the universe. The  expression (\ref{lineargravity}) fulfils our basic physical requirement on the action that it should has the dimension of length and should reduce to the action of the relativistic particle in the situations when a manifold collapses to a one-dimensional 
curve  
\be\label{relatparticle}
L=-M c \int  d s.
\ee
Both expressions contain the geometrical invariants which are in general not positive-definite under the square root. In the relativistic particle case (\ref{relatparticle}) the expression under the root becomes negative for a particle moving with a velocity which exceeds the velocity of light. In that case the action develops an imaginary part and quantum-mechanical suppression of amplitudes prevents a particle from exceeding the velocity of light \cite{Pauli:1941zz,Feynman:1949hz,Feynman:1949zx}. 
A similar mechanism  was implemented in the Born-Infeld modification of electrodynamics with the aim to prevent the appearance of infinite electric 
fields \cite{borninfeld,Nielsen:1973qs,Guendelman:2011sm}.  It was found that there is a deep relation between the maximum field strength action and the fact that the D0-brane velocity is limited by the velocity of light \cite{Bachas:1995kx}.  The idea of a limiting curvature action was developed in the  articles \cite{Markov:1982ed,Brandenberger:1993ef,Chamseddine:2016ktu,Chamseddine:2016uef} to prevent gravitational singularities. 
 
 One can expect that 
in the case of modified gravity with linear action $L$ in (\ref{lineargravity}) there may appear space-time regions which are unreachable by the test particles
if in that regions the expression under the root becomes negative. If these "locked"  space-time regions happen  to appear and if that space-time regions 
include singularities, then one can expect  
that the gravitational singularities are naturally excluded from the theory  due to the fundamental principles of quantum mechanics. The question of consistency  of the new action principle, if it is  the 
right one, can only be decided by their physical consequences. 

In the next sections we shall consider the  black hole singularities and the physical effects which are induced  by the inclusion of the linear action (\ref{lineargravity}). As
we shall see, the expression under the root becomes negative in the region which is smaller than the Schwarzschild radius $r_g$ and includes the singularities. For the observer which is far away from the horizon the linear action perturbation induces a small additional advance precession of the perihelion, but has a profound influence on the physics behind  the horizon. We are confronted here with a drastically unusual concept that there may exist 
space-time regions which are excluded from the physical scene, being physically unreachable by test particles or observables. If one accepts 
this concept, then it seems plausible that the gravitational singularities are  excluded from the modified theory. In this paper we have only taken the first steps to describe the phenomena  which are caused by the additional linear term in the gravitational action proposed in \cite{Savvidy:1995mr,Ambjorn:1996kk,Savvidy:1997qf}. 

  \section{\it  Schwarzschild Black Hole Singularities}

The modified action which we shall consider is a sum 
\beqa\label{actionlineargravity23}
 && S =-{  c^3\over 16 \pi G  } \int R \sqrt{-g} d^4x 
-  M c  \int {3  \over 8 \pi}(1-\gamma)\sqrt{ I_1+ \gamma I_2
}~\sqrt{-g }  d^4x,
\eeqa
where we introduced a dimensionless parameter $\gamma$ in order to consider a general linear combination of the invariants. 
The additional linear term has high derivatives of the metric and the equations of motion which follow from the variation of the action are much more complicated than in the standard gravity case, but because the linear action became relevant only in the situations when the metric is changing relatively fast one can consider as a first approximation the perturbation of the solutions of standard gravity generated by the additional linear term. It is obvious that {\it in the space-time regions where the metric is varying slowly the modification of the standard gravity solutions should  be  negligible, at the same time the perturbations  may became relevant in the space-time regions where curvature is large and the metric is changing relatively fast.}

\begin{figure}
\begin{center}
\includegraphics[width=6cm]{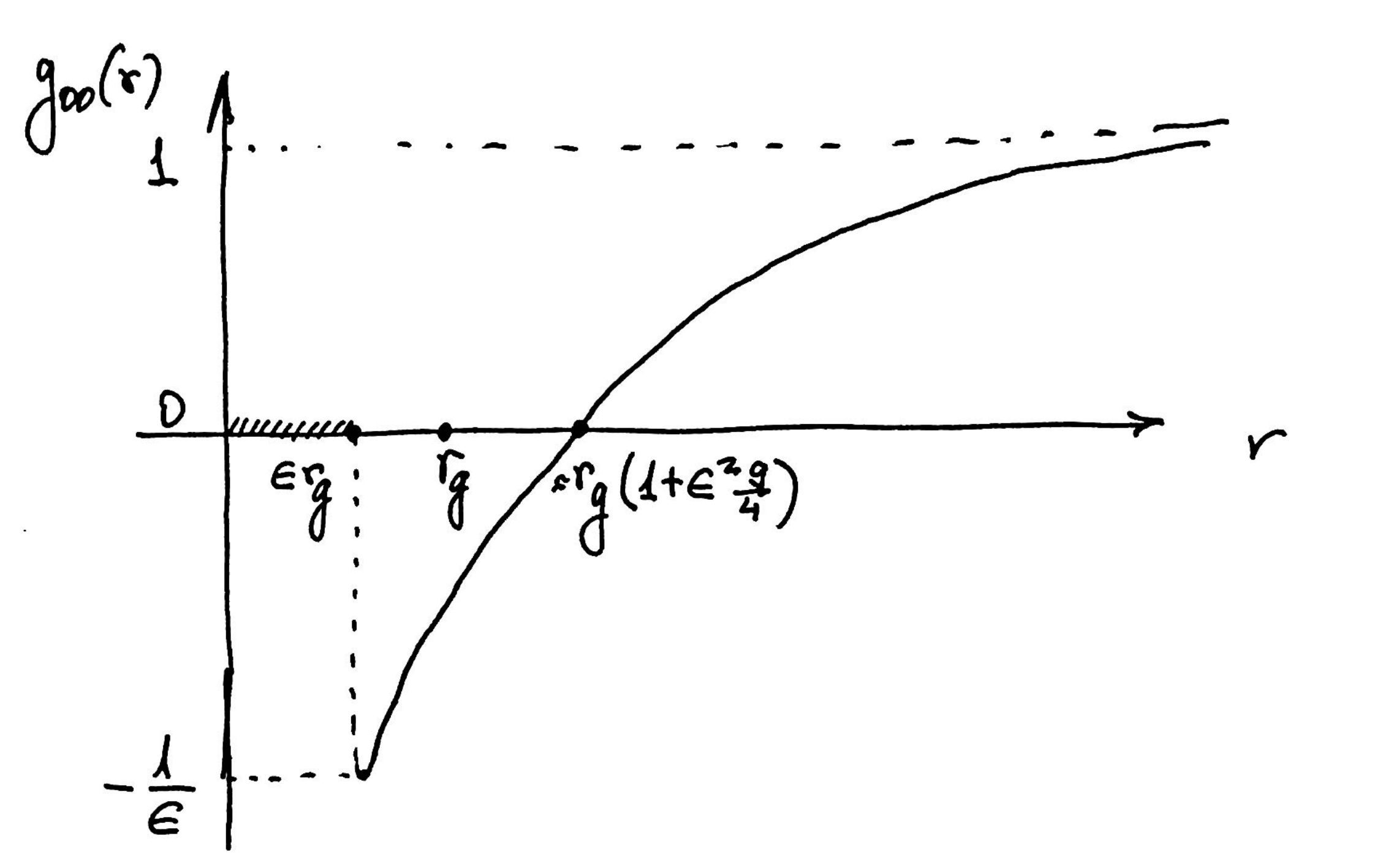}~~~~~~~~~~
\caption{
The graphic of the potential function $g_{00}=1-{r_g\over r} -  \Big[1-
\Big(1 -\varepsilon  {r_g\over r} \Big)^{3/2} \Big]^2 $. 
At $r \rightarrow \infty$ the 
$g_{00} \rightarrow 1$ and at $r \rightarrow \varepsilon r_g $ the 
$g_{00} \rightarrow -{1\over \varepsilon}  $.  
}
\label{fig5}
\end{center}
\end{figure}

In this section we shall consider the perturbation of the Schwarzschild solution which is induced  by the the additional linear term in the action and try to understand how it influences the black hole physics and the singularities. The Schwarzschild solution has the form 
\be\label{schwarz}
ds^2 = (1-{r_g\over r}) c^2 dt^2 - (1-{r_g\over r})^{-1} dr^2 -r^2 d\Omega^2~,
\ee
where 
$
g_{00} =1-{r_g\over r},~g_{11} =-(1-{r_g\over r})^{-1},~g_{22} =-r^2,~g_{33} =-r^2 \sin^2\theta,
$
and 
$$
r_g = {2GM \over c^2},~~~~ \sqrt{-g} = r^2 \sin\theta.
$$
The nontrivial quadratic curvature invariant in this case has the form
\beqa
&&I_0={1\over 12} R_{\mu\nu\lambda\rho}  R^{\mu\nu\lambda\rho} = ({r_g \over r^3})^2  \eeqa
 and shows that the singularity located at $r=0$ is actually a curvature singularity.
 The event horizon is located where the metric component $g_{rr}$ diverges, that is,  at 
$
r_{horizon}=  r_g . 
$
The expressions for the two curvature polynomials (\ref{lineargravity2}) of our interest are\footnote{It should be stressed that all other invariant polynomials of the same dimensionality can be expressed in terms of $I_1$ and $I_2$, and they are given in Appendix.}:
\beqa
I_1= {r^2_g( r -  r_g )  \over  r^9},~~~~~I_2={r^3_g \over  r^9}~, 
\eeqa
and on the Schwarzschild solution the action acquires additional term of the form 
\beqa\label{lineargravityS}
L =  -M c^2    \int  {3  \over 2} \varepsilon  \sqrt{  1-\varepsilon {  r_g   \over  r }}~ {r_g \over r^2} dr dt~,
\eeqa
where
\be
\varepsilon = 1-\gamma . \nn
\ee
As one can see, the expression under the square root in (\ref{lineargravityS}) becomes negative at
\be 
r < \varepsilon r_g, ~~~~~0 < \varepsilon \leq 1
\ee
and defines  the region in which the action develops an imaginary part.  Using the analogy with the relativistic particle action (\ref{relatparticle})
\be\label{relatparticle1}
L=-M c \int  d s = -M c^2 \int \sqrt{1-{\vec{v}^2\over c^2 }}~ dt.
\ee
were the imaginary part is developing at the velocities which are larger than the velocity of light. In 
the last case this leads to the destructive superposition of quantum mechanical amplitudes 
outside the light-cone forbidding a particle to move with $v >c$. 
One can expect that in our case as well the development of the imaginary action in (\ref{lineargravityS})  will lead to 
region of the space-time which is unreachable by the test particles.  The size of the region  depends on the parameter $\varepsilon$ and is smaller than the gravitational radius $r_g$   if $\varepsilon$ is less than one (see Fig. \ref{fig5}). 

This result seems to have profound consequences on the gravitational singularity at $r=0$. 
In a standard interpretation of the singularities, which appear in spherically symmetric gravitational collapse,  the singularity at $r=0$ is hidden in the sense that no signal from it can reach infinity. The singularities
are not visible for the outside observer, but hidden behind an event horizon. In that interpretation the singularities are still present in the theory. In the suggested 
scenario it seems possible to eliminate  the singularities from the theory based on the fundamental principles of quantum mechanics.   The singularities are excluded from the theory on the same level as the motion of particles with a velocity which
exceeds the speed of light. 

The quantum mechanical amplitude in terms of the path integral 
has the form 
\be
\Psi = \int e^{{i \over  \hbar } S[g]} \CD g_{\mu\nu}(x) \nn,
\ee 
where integration is over all diffeomorphism nonequivalent metrics.
For the Schwarzschild massive object which is at rest we can find the expression for the action
integrating (\ref{lineargravityS}) 
\be
L = - M c^2    \int^{\infty}_{\varepsilon r_g} {3  \over 2} \varepsilon  
\sqrt{  1-\varepsilon {  r_g   \over  r }}~ {r_g \over r^2} dr dt \\
=M c^2   t   
\ee
and confirm that it is proportional to the length $t$  of the space-time trajectory, as it should be for the relativistic particle  at rest, 
so that the corresponding amplitude can be written in the form
\be
\Psi \approx  \exp{ ({i   \over   \hbar } \sum_n M c^2 t)}. 
\ee

The perturbation (\ref{lineargravityS}) generates a contribution to the distance invariant $ds$ in (\ref{schwarz}) of the form
\be\label{hatper}
{3  \over 2}  \int^{\infty}_{r} \varepsilon \sqrt{  1-\varepsilon {  r_g   \over  r }}~ {r_g \over r^2} dr =    \Big[1-
\Big(1 -\varepsilon  {r_g\over r} \Big)^{3/2} \Big]
\ee
and allows to calculate  the correction to the purely temporal component of the metric tensor (\ref{schwarz})  caused by the additional term in the  action
\beqa\label{main}
&&g_{00} =1-{r_g\over r} -    \Big[1-
\Big(1 -\varepsilon  {r_g\over r} \Big)^{3/2} \Big]^2  .
\eeqa
Using the above expression of the metric one can analyse the influence of the perturbation 
on the physics at different regions of the space-time. 
The  equation used to determine gravitational time dilation near a massive body is modified in this case and the  proper time between events is defined now by the equation
\be
d \tau = \sqrt{g_{00}} dt = \sqrt{1-{r_g\over r} -    \Big[1-
\Big(1 -\varepsilon  {r_g\over r} \Big)^{3/2} \Big]^2} dt
\ee
and therefore
$
d \tau   \leq dt,
$ as in standard gravity. 
It follows from (\ref{main}) that near the gravitational radius $r \approx r_g$ a purely temporal component of the metric tensor has the form 
\beqa
g_{00} ~\approx ~1-{r_g\over r} - \varepsilon^2 {9 \over 4}   \Big( {r_g\over r} \Big)^{2} 
 +  \CO( \varepsilon^3)
\eeqa
confirming that the perturbation is small and the infinite red shift which appears in the standard case at $r=r_g$ now appears in its small vicinity
\be
r \approx  ~ r_g (1+ {9 \over 4} \varepsilon^2)  +\CO(\varepsilon^4) .
\ee 
To define the perturbation of the  trajectories of the test particles outside of the massive body we shall study the behaviour  of the solutions of the Hamilton-Jacobi equation for geodesics, which is modified by the perturbation of the metric: 
\beqa
&g^{\mu\nu} {\partial A \over \partial x^{\mu}}{\partial A \over \partial x^{\nu}}~
  = g^{00}  \Big({\partial A \over c \partial t}\Big)^2-
{1\over g^{00}} \Big({\partial A \over  \partial r}\Big)^2 - {1\over r^2}\Big({\partial A \over  \partial \phi}\Big)^2 =m^2 c^2.\nn
\eeqa
The solution has the form
\be
A = -E t + l \phi +A(r),
\ee  
where  $E$ and $l $ are the energy and angular momentum of the test particle and  
\be\label{geodesic}
A(r)= \int \Big[ \Big( g^{00} {E^2 \over c^2} - m^2 c^2 -{l^2 \over r^2} \Big) g^{00} \Big]^{1/2} dr .
\ee
In the non-relativistic limit $E=E^{'} + m c^2, E^{'} \ll m c^2$, and in terms of a new coordinate 
$r(r-r_g)=r^{'}$
we shall get 
\be
A(r) \approx \int \Big[   ( {E^{'2} \over c^2}  +2 E^{'} m) 
+{1 \over r^{'}} (4 E^{'} m r_g + m^2 c^2 r_g) 
- {1 \over r^{' 2}} \Big(  l^2 - {3  \over 2} m^2 c^2 r^2_g (1 + {3  \over 2}  \varepsilon^2)    \Big) \Big]^{1/2} dr^{'} .
\ee
The geodesic trajectories are defied by the equation 
$  \phi + \partial  A(r) / \partial l = Const$  and the advance precession of the 
perihelion $\delta \phi$  expressed in radians per revolution is  given by the expression 
\be
\delta \phi =  {3 \pi m^2 c^2 r^2_g  \over 2 l^2 } (1 + {3  \over 2}  \varepsilon^2)=
{6 \pi G M   \over c^2 a (1-e^2) } (1 + {3  \over 2}  \varepsilon^2),
\ee
where $a$ is the semi-major axis and $e$ is the orbital eccentricity. As one can see from the above result, the precession is advanced by the additional factor $1 +{3  \over 2}  \varepsilon^2$. The upper bound on the value of  $\varepsilon$ can be extracted from the observational data for the advanced precession of the Mercury perihelion,  which is $42,98 \pm 0,04$ seconds of arc per century, thus
$$
\varepsilon \leq 0,16~.
$$
For the light propagation we shall take $m^2 =0$,~ $E= \omega_0$,~$l =  \rho~  \omega_0 / c$ in (\ref{geodesic}):
\be
A(r)=  {\omega_0 \over c} \int \sqrt{  ( g^{00}   - {\rho^2    \over   r^2} ) g^{00} }  dr  \approx   {\omega_0 \over c} \int \sqrt{ 1+ 2 {r_g\over r} - {\rho^2    \over   r^2}   }  dr +\CO( \varepsilon^2  r^2_g /r^2). 
 \ee
The trajectory is defined by the equation $ \phi + \partial  A(r) / \partial \rho = Const$ and in the given approximation the deflection of light  ray remains unchanged:
\be
\delta \phi =  2 {r_g\over \rho},
\ee
where $\rho$ is the distance from the centre of gravity. The deflection angle is not influenced by the perturbation, which is of order $\CO( \varepsilon^2  r^2_g / \rho^2)$,  and does not impose a sensible constraint on $\varepsilon$. 
In the next sections we shall consider perturbation of the Reissner-Nordstr\"om and the Kerr solutions.

\section{\it  Reissner-Nordstr\"om Solution}

The Reissner-Nordstr\"om solution has the form 
\be\label{reissnernordstrom}
ds^2 = (1-{r_g\over r} + {r^2_{Q}\over r^2}) c^2 dt^2 - (1-{r_g\over r} + {r^2_{Q}\over r^2})^{-1} dr^2 -r^2 d\Omega^2~,
\ee
where 
$$
r_g = {2GM \over c^2},~~~~r^2_{Q} = { Q^2 G \over c^4},~~~~~~\sqrt{-g} = r^2 \sin\theta.
$$
The nontrivial quadratic curvature invariant is
\beqa
&&I_0={1\over 12} R_{\mu\nu\lambda\rho}  R^{\mu\nu\lambda\rho} = {3 r^2 r^2_g  -12 r r_g r^2_Q +14 r^4_Q\over 3 r^8},  
\eeqa
 and it shows that the singularity is located at $r=0$.
The event horizon and internal Cauchy horizon are located where the metric component $g_{rr}$ diverges: 
\be
r_{\pm}= {1\over 2}(r_g \pm \sqrt{r_g^2 -4 r^2_Q}).
\ee
The solutions with $r_Q > r_g/2$ represent a naked singularity.  As we shall see below, at these charges $r_Q$ the linear action develops a complex value  and  prevents the appearance of the naked singularities.

The expression for the two curvature polynomials of our interest are:\beqa
&&I_1= {  (r^2 - r r_g + r_Q^2) (45 r^2 r_g^2 - 216 r r_g r_Q^2 + 304 r_Q^4) \over 45r^{12}  },\nn\\
&&I_2={9 r^3 r_g^3 + 36 r^3 r_g r_Q^2 - 96 r^2 r_g^2 r_Q^2 - 88 r^2 r_Q^4 + 
 264 r r_g r_Q^4 - 200 r_Q^6 \over 9 r^{12}}. 
\eeqa
It is convenient to introduce the  dimensionless quantities:
\be\label{units}
\hat{r} = {r \over r_g},~~~~~\hat{r_Q}= {r_Q \over r_g},
\ee
and express the linear action on the Reissner-Nordstr\"om solution in the following  form: 
\beqa\label{lineargravityRN}
&&L = - M c^2    \int   { 3  \over 2    } \varepsilon \sqrt{  \Big(1    -  {b \over \hat{r}}  
- {c \over \hat{r}^2 }  
+ {f \over \hat{r}^3}   - {e \over \hat{r}^4 } \Big)}   ~ {1  \over \hat{r}^2 } d \hat{r}    dt ~,
\eeqa
where 
\beqa
&& b=   \varepsilon +  \hat{r}_Q^2 { 4  (1 +5 \varepsilon) \over 5  },~~~
c=   \hat{r}_Q^2 {(219 - 480 \varepsilon) \over 45} 
+ \hat{r}_Q^4 {8 (17 - 55 \varepsilon) \over 45 },\nn\\ 
&&~~~
f=  \hat{r}_Q^4 {8  (20 - 33 \varepsilon) \over 9 }, ~~~e= \hat{r}_Q^6 {8 (87 - 125\varepsilon) \over 45 }.\nn
\eeqa
If the charge of the black hole is equal to zero, $ \hat{r}_Q=0$, then the action (\ref{lineargravityRN}) reduces to the expression (\ref{lineargravityS}) on the Schwarzschild solution.  
For  the extremal black hole of the change $r_Q =r_g/2$  the  fourth order polynomial under the root in (\ref{lineargravityRN}) is positive for  $r >   r_g/2 $, is equal to zero at $r =  r_g/2  $ and is negative for $r <  r_g/2 $. The  region which is "locked" for the test particles in this case is defined by $r =  r_g/2  $ and  
prevents the appearance of the naked singularities.  

It is helpful to represent the polynomial under the square root in (\ref{lineargravityRN}) in the form 
\be
\Big(1    -  {b \over \hat{r}}  
- {c \over \hat{r}^2 }  
+ {f \over \hat{r}^3}   - {e \over \hat{r}^4 } \Big)= (1- {\hat{r}_1 \over \hat{r}})...(1-{\hat{r}_4 \over \hat{r}})~,
\ee
where $\hat{r}_i,~i=1,...,4$ are the roots of the fourth order polynomial.
The largest positive real valued root  at which the polynomial turns out to be negative is defined as $\hat{r}_4$. Near that radius one
can approximate the polynomial as 
\be
(1- {\hat{r}_1 \over \hat{r}_4})(1- {\hat{r}_2 \over \hat{r}_4})(1- {\hat{r}_3 \over \hat{r}_4})~(1-{\hat{r}_4 \over \hat{r}}) = \Gamma(\hat{r}_1,...,\hat{r}_4) ~(1-{\hat{r}_4 \over \hat{r}}).
\ee
Thus the action will take the form
\beqa\label{lineargravityRN1}
&&S\approx  - M c^2  \Gamma  \int   { 3  \over 2    } \varepsilon \sqrt{   (1-{\hat{r}_4 \over \hat{r}})}   ~ {1  \over \hat{r}^2 } d \hat{r}    dt 
\eeqa
and it has a form similar to the case we had in the Schwarzschild  black hole  (\ref{lineargravityS}). As one can see, the expression under the square root in (\ref{lineargravityRN1}) becomes negative at
\be 
r <  r_4(r_Q,\varepsilon)
\ee
and defines  the region which is unreachable by the test particles.

The Table \ref{tbl:largeN} presents the values of the radius $\hat{r}_4$ at which the polynomial under the root changes its sign from positive to negative and the action becomes complex as a function of the 
charge $\hat{r}_Q$ and the parameter $\varepsilon$.  As it follows from the Table \ref{tbl:largeN} for the extremal black hole, $\hat{r}_Q  =1/2$,  the "locked" region has the radius  $\hat{r}_4 =  1/2 $  and increases with the charge $\hat{r}_Q >1/2$,  preventing the appearance of naked singularities. At  $\hat{r}_Q =1$ the locked region has the radius $\hat{r}_4 \approx 2.14$. For $\hat{r}_Q  < 1/2$ the locked region is smaller than horizon 
 $\hat{r}_4 < 1/2$.
 \begin{table}[htbp]
   \centering
   \begin{tabular}{@{} lcccrcl @{}}        \toprule
      Charge & Parameter  & Maximal real solution  \\
      $\hat{r}_Q$    & $ \varepsilon$ &  $\hat{r}_4$    \\  
      \midrule
        1   & 0.1 & 2.14   \\
	1/2 & 0.1 & 0.50   \\
        1/4& 0.1 & 0.47   \\
 	1/8 & 0.1 &0.29   \\
	1/16 & 0.1 &0.18    \\
      \bottomrule
   \end{tabular}
   \caption{ The table of  the solutions $\hat{r}_4$ at which the four order polynomial under the root function in (\ref{lineargravityRN})  becomes negative.  The value of $\hat{r}_4$ is measured  in $r_g$ units  (\ref{units}) . For the extremal black hole $\hat{r}_Q  =1/2$  the "locked" region has the radius  $\hat{r}_4 =  1/2 $  and increases with the black hole charge $\hat{r}_Q >1/2$, thus  preventing the appearance of naked singularities. For $\hat{r}_Q  < 1/2$ the locked region is smaller than horizon 
 $\hat{r}_4 < 1/2$, that is $r_4 <  r_g / 2$.}
   \label{tbl:largeN}
\end{table}

 \section{ \it Kerr  Solution  }

Let us also consider the Kerr metric  
\be
ds^2 = (1-{r_g r \over \rho^2}) c^2 dt^2 - {\rho^2 \over r^2 - r_g r +a^2} dr^2 -\rho^2 d\theta^2
-(r^2 +a^2 + {r_g r a^2 \over \rho^2 }\sin^2\theta)\sin^2\theta d\phi^2  + 2 {r_g r a \over \rho^2}\sin^2\theta d\phi cdt~,
\ee
where 
\beqa
&g_{00} = 1-{r_g r \over \rho^2},~g_{11} =- {\rho^2 \over r^2 - r_g r +a^2},~g_{22} =-\rho^2,~
g_{33} =-(r^2 +a^2 + {r_g r a^2 \over \rho^2}\sin^2\theta)\sin^2\theta,\nn\\
&g_{03}= g_{30}= {r_g r a \over \rho^2}\sin^2\theta 
\eeqa
and 
$$
a = {J\over Mc },~~~r_g = {2GM \over c^2},~~~~ \rho^2 = r^2 +a^2 \cos^2\theta, ~~~~\sqrt{-g} = \rho^2 \sin\theta.
$$
The nontrivial quadratic curvature invariant is
\beqa
&&I_0={1\over 12} R_{\mu\nu\lambda\rho}  R^{\mu\nu\lambda\rho} = 
{r^2_g(r^2-a^2 \cos^2\theta) [ (r^2 + a^2 \cos^2\theta)^2 -16 a^2 r^2 \cos^2\theta] \over  (r^2 + a^2 \cos^2\theta)^6} 
\eeqa
and it shows that the singularity located at 
$
r=0,~~~\theta = \pi/2
$
is a curvature singularity.
The event horizon is defined by the largest root of the equation 
$r^2 - r_g r +a^2=0$ where the metric component $g_{11}$ diverges:    
\be\label{horiz}
r_{horizon}= {1\over 2}(r_g + \sqrt{r^2_g-4 a^2}).
\ee
For $a > r_g/2$  there are no  real valued  solutions and there is no event horizon. With no event horizons to hide it from the rest of the universe, the black hole ceases to be a black hole and will instead be a naked singularity.
The outer ergosurface is defined by the equation where the purely temporal component $g_{00}$ of the metric changes the sign from positive to negative:
\be
r_{ergosur}= {1\over 2}(r_g + \sqrt{r^2_g-4 a^2 \cos^2\theta}).
\ee
These two critical surfaces are tangent to each other at poles $\theta = 0,\pi$ and they exist only when $a < r_g/2$.  The space between these two surfaces defines the ergosphere. At maximum value of the angular momentum $a = r_g/2$ these surfaces are defined by the equations
\be\label{maxang}
r_{horizon}= {r_g\over 2}~,~~~r_{ergosur}= {r_g\over 2}(1 + \sin \theta).
\ee
Let us now consider the expressions for the curvature polynomials 
$I_1$ and $I_2$ in the case of Kerr solution
\beqa
&&I_1=  
 {r^2_g(r^2- r_g r + a^2 \cos^2\theta) [ r^8 -28 a^2 r^6 \cos^2\theta +70 a^4 r^4 \cos^4\theta - 28 a^6 r^2 \cos^6\theta + a^8  \cos^8\theta] \over  (r^2 + a^2 \cos^2\theta)^9},\nn\\
&&I_2  = {r r^3_g (r^2 - 3 a^2 \cos^2\theta) (r^6 - 33 a^2 r^4 \cos^2\theta 
+27 a^4 r^2 \cos^4\theta - 
   3 a^6 \cos^6\theta) ) \over (r^2 + a^2 \cos^2\theta )^9 }.
\eeqa
It is convenient to introduce the  dimensionless quantities 
\be\label{unitsk}
\hat{r} = {r \over r_g},~~~~~\hat{a}= {a \over r_g},~~~~~\sigma^2= \hat{a}^2 \cos^2\theta
\ee
so that the linear action will takes the form
\beqa\label{lineargravityK}
&&S = - M c^2    \int   { 3  \over 2    } \varepsilon \sqrt{  f(\hat{r}, \hat{a},\varepsilon,\theta) }   ~ {1  \over \hat{r}^2 } d \hat{r}    dt ~,
\eeqa
where
\beqa\label{poly}
&&f(\hat{r}, \hat{a}, \varepsilon,\theta ) = \Big(1 - {\varepsilon \over \hat{r}} - {27   \sigma^2 \over \hat{r}^2} - {8  \sigma^2 \over \hat{r}^3} 
+ { 36  \sigma^2 \varepsilon \over \hat{r}^3} + {42   \sigma^4 \over \hat{r}^4} + {56   \sigma^4 \over \hat{r}^5} 
- { 126   \sigma^4 \varepsilon \over \hat{r}^5} +\nn\\
&&+ {42  \sigma^6 \over \hat{r}^6} - {56   \sigma^6 \over \hat{r}^7} 
+ {84   \sigma^6 \varepsilon \over \hat{r}^7} - {27   \sigma^8 \over \hat{r}^8} + {8   \sigma^8 \over \hat{r}^9} 
- { 9   \sigma^8 \varepsilon \over \hat{r}^9} + { \sigma^{10} \over \hat{r}^{10}} \Big) 
\Big(1+ {   \sigma^2 \over \hat{r}^2} \Big)^{-9} .
\eeqa
The region which is locked for the test particles is defined by the largest real positive root of the polynomial (\ref{poly}) at which the polynomial turns out to be negative. It is  denoted as $\hat{r}_{10}=\hat{r}_{10}(\hat{a}, \theta,\varepsilon)$.
Let us consider the situation with maximal angular momentum $a=r_g/2$ as in (\ref{maxang}).
 The roots $\hat{r}_{10}(1/2, \theta,\varepsilon,)$ can be found numerically for different parameters of the Kerr solutions.  In the maximal angular momentum case some of the values are: 
\beqa
&&a=r_g/2,~~ \theta = 0,\pi ~~~~ r_{horizon}= r_g/2,~~ r_{ergosur}= r_g/2,~~
r_{10} = 2.6 r_g \nn\\
&&a=r_g/2,~~ \theta = \pi/2, ~~r_{horizon}= r_g/2,~~ r_{ergosur}= r_g,~~
r_{10} =   r_g /2 .
\eeqa
Thus the singularity is unreachable by the test particles. In the case of smaller angular momentum the locked region shrinks and is inside the event horizon.

In the above sections we were considered the perturbation of the exact solutions of the classical gravity by the linear action which suppresses the singular fluctuations. It is a difficult task to find the exact solutions of the equations which follow from the action (\ref{lineargravity}) and (\ref{actionlineargravity23}) and we were unable to find exact solutions of these equations, but in the first approximation consider above the results are pointing out into the existence of the space-time regions surrounding the singularity which are inaccessible by the test particles.  

In conclusion I  would like to thank Jan Ambjorn for invitation  and
kind hospitality in the Niels Bohr Institute, where part of the work was done. 
I  would like to thank Alex Kehagias for references and Kyriakos Papadodimas for useful 
remarks. The author acknowledges support by the ERC-Advance Grant 291092,  "Exploring the Quantum Universe" (EQU) and COST Action MP1405.  This lecture was presented in the Corfu Summer Institute 2017 "School and Workshops on Elementary Particle Physics and Gravity", 2-28 September 2017, Corfu, Greece and COST Action MP1405   "Quantum Structure of Spacetime"
III.  Annual Workshop: Quantum Spacetime '18, 19 - 23 February 2018,  Sofia, Bulgaria. 
 
\section{ \it Appendix}

The general form of the linear action has the form: 
\beqa\label{generalactionlineargravity}
 && S_L = 
-   m_P c  \int {3  \over 8 \pi}\sqrt{\sum^{3}_{1}  \eta_i K_i+\sum^{4}_{1}  \chi_i J_i+  \sum^{9}_{1}  \gamma_i I_i}~\sqrt{-g }  d^4x ~,\nn
\eeqa
where the curvature invariants have the form 
\beqa
&&I_0={1\over 12} R_{\mu\nu\lambda\rho}  R^{\mu\nu\lambda\rho},    \nn\\
&& I_1= -{1\over 180}R_{\mu\nu\lambda\rho;\sigma} R^{\mu\nu\lambda\rho;\sigma} ~,\nn\\
&&I_2=+{1\over 36} R_{\mu\nu\lambda\rho}  \Box  R^{\mu\nu\lambda\rho}~,\nn\\
&&I_3= -{1\over 72} \Box (R_{\mu\nu\lambda\rho} R^{\mu\nu\lambda\rho} )=
5 I_1 -I_2 ~, \nn\\
&&I_4= -{1\over 90} R_{\mu\nu\lambda\rho;\alpha} R^{\alpha\nu\lambda\rho;\mu} =   I_1 ~,\nn\\
&&I_5=  - {1\over 18}  (R^{\alpha\nu\lambda\rho}  R^{\mu}_{~~\nu\lambda\rho})_{;\mu;\alpha}   =  5 I_1 -I_2~, \nn\\
&&I_6=  -{1\over 18}  (R^{\alpha\nu\lambda\rho}  R^{\mu}_{~~\nu\lambda\rho})_{;\alpha;\mu}  = 5 I_1 -I_2~, \nn\\
&&I_7=  {1\over 18} R^{\alpha\nu\lambda\rho} R^{\mu}_{~\nu\lambda\rho;\alpha;\mu}  =  I_2 ~,\nn\\
&&I_8= R^{\mu}_{~\nu\lambda\rho;\mu} R^{\sigma\nu\lambda\rho}_{~~~~~;\sigma}~,  \nn\\
&&I_9=  R^{\alpha\nu\lambda\rho} R^{\mu}_{~\nu\lambda\rho;\mu;\alpha} ~,\nn\\
&&J_0=  R_{\mu\nu }  R^{\mu\nu } ~,~~J_1= R_{\mu \nu;\lambda} R^{\mu\nu;\lambda }~,~~J_2=  R^{\mu\nu }   \Box R_{\mu\nu }~,~~J_3=  \Box (R^{\mu\nu }    R_{\mu\nu })~,~~J_4=   R_{\mu \sigma}^{~~~;\mu} R^{\nu\sigma }_{~~~;\nu} \nn\\
&& K_0=  R^2~,~~
 K_1=  R_{;\mu}  R^{;\mu}~,~~
 K_2=  R    \Box R~,~~
 K_3=  \Box  R^2~  . 
\eeqa
The $\eta_i , \chi_i $ and $ \gamma_i$ are free parameters. Some of the invariants can be expressed through others using Bianchi identities.

\end{document}